\documentclass[prd,superscriptaddress,twocolumn]{revtex4}

\usepackage{caption}
\usepackage{subcaption}
\usepackage{amssymb}
\usepackage{amsfonts}
\usepackage{amsmath}
\usepackage{graphicx}
\usepackage{dcolumn}
\usepackage{bm}

\begin{document}
\title{Chronology protection in $f(T)$ gravity: the case of Gott's pair of moving cosmic strings}

\author{Franco Fiorini}
\email{francof@cab.cnea.gov.ar} \affiliation{Departamento de Ingenier\'{i}a en Telecomunicaciones, Consejo Nacional de Investigaciones Cient\'{i}ficas y T\'{e}cnicas (CONICET) and Instituto Balseiro (UNCUYO), Centro At\'{o}mico Bariloche, Av. Ezequiel Bustillo 9500, CP8400, S. C. de Bariloche, Rio Negro, Argentina.}

\begin{abstract}
As a consequence of the spacetime structure, defined by the tetrad field instead of the metric tensor alone, $f(T)$ gravity seems to harbor its own chronology protection agency. When Gott's pair of moving cosmic strings is considered, it is shown that the requirement of having a global parallelization --i.e. a global smooth field of tetrads-- drastically restrict the form of the tetrads on the junction surface between the two strings. The junction conditions on the tetrad field are satisfied only if the corresponding boosts needed to put the strings in motion are null on the matching surface. This seems to throw overboard Gott's construction from the outset without the need of analyzing the divergence of the expectation value of the energy momentum tensor on the Cauchy horizon, evading in this way bothering quarrels concerning the choice of vacuum. 
\end{abstract}
\maketitle

\section{Introduction}
It is well known that General Relativity (GR) allows the possibility of time travel. Among the various solutions presenting causal anomalies of one sort or another, Gott's spacetime is perhaps the most peculiar \cite{Gott1}. Being a vacuum solution of Einstein's equations, the closed timelike curves (CTC's) arising in Gott's solution are not powered by any form of matter-energy. Furthermore, the two-moving strings spacetime is constructed by pasting two copies of the conical geometry characterizing the exterior metric of an infinitely long cosmic string, so it differs from Minkowski space only in its global properties. Additionally, the spacetime so obtained is geodesically complete, reason why the non-compact region containing causal violations is not affected by fresh information coming from curvature singularities. Finally, the Cauchy horizon in Gott's solution is not compactly generated, which makes Hawking's chronology protection conjecture inapplicable \cite{Haw}. For a detailed, technical description of time travel in GR, we can refer the reader to \cite{timemachines} and \cite{timemachines2}.

On physical grounds, Gott's construction suffers from two drawbacks: first, cosmic strings were not yet detected, a fact that should not worry us too much, specially if we think on the immeasurable advancements achieved in regard with the detection of objects as elusive as black holes and gravitational waves. It is by no means an exaggeration to conceive the possibility of discovering not far ahead in the future topological defects on large scales, as cosmic strings \cite{CStrings}. Second, if any, cosmic strings are quite likely finite in length, or come in loops, in which case Hawking's chronology protection conjecture would enter into the scene. However, the expectation value of the stress-energy tensor on the Cauchy horizon is strongly dependent on the choice of vacuum, as was explicitly shown in \cite{Li} and \cite{Gott3} for Misner space. This result is very important because Gott's two-strings space can be thought as a generalization of Misner space \cite{Grant}. Hence, it seems unclever to judge by means of this argument whether or not CTC's can survive the perturbations that quantum fields would produce on the metric. Additional physical arguments against the production of CTC's by cosmic strings were considered in \cite{Deser}. 

It is our intention here to discuss Gott's construction in the context of $f(T)$ gravity. In order to present a relatively self-contained exposition, we revisit in section \ref{sec2} the solutions describing both, the interior and exterior structures associated to infinitely long cosmic strings in GR, followed by a concise review of Gott's two-moving strings space. Section \ref{sec3a} deals with the interior and exterior tetrads describing the corresponding cosmic strings solutions according to $f(T)$ gravity, as well as the junction conditions involved in their matching. Section \ref{sec3b} is entirely devoted to the treatment of the two-moving strings space, and to show that no traveller is able to circumnavigate both strings with the aim of visiting his/her own past. 

\section{Infinitely long cosmic strings in GR and Gott's construction}\label{sec2}

When dealing with just one string, it is convenient to write down the metric in cylindrical coordinates.
The interior metric is (we follow \cite{Gott2} in the first part of this section)
\begin{equation}\label{int}
ds_{-}^2=-dt^2+r_{0}^2(d\theta^2+\sin^2\theta\, d\varphi^2)+dz^2,
\end{equation}
which is a solution of Einstein's field equations $R_{\mu\nu}-\frac{1}{2}\,R\, g_{\mu\nu}=8 \pi T_{\mu\nu}$, for an energy-momentum tensor $T_\mu{}^{\nu}=diag(\rho_{0},0,0,-p_{z})$, with $p_{z}=-\rho_{0}$ and $\rho_{0}=1/8\pi r_{0}^2$. The coordinates in (\ref{int}) range according to $-\infty<t,z<\infty$, $0\leq\varphi<2\pi$ and $0\leq\theta\leq\theta_{M}$. A change of coordinates to $r=r_{0}\,\theta$ give us
\begin{equation}\label{int2}
ds^2_{-}=-dt^2+dr^2+r_{0}^2\sin^2(r/r_{0}) d\varphi^2+dz^2,
\end{equation}
which will be useful in brief. The exterior, vacuum metric, in turn reads
\begin{equation}\label{ext1}
ds^2_{+}=-dt^2+dr^2+(1-4\mu)^2r^2d\varphi^2+dz^2,
\end{equation}
where $\mu$ is the mass per unit $z$-length of the string, namely
\begin{equation}\label{mass}
\mu=\rho_{0}\, r_{0}^2\int_{0}^{2\pi}\int_{0}^{\theta_{M}}\sin\theta \, d\theta\, d\varphi=(1-\cos\theta_{M})/4,
\end{equation}
where we have used that $\rho=\rho_{0}=1/8\pi r_{0}^2$. It is clear that (\ref{ext1}) can be cast in the form
\begin{equation}\label{ext2}
ds^2_{+}=-dt^2+dr^2+r^2d\varphi'^{\,2}+dz^2,
\end{equation}
provided $\varphi'=(1-4\mu)\varphi$, then $0\leq\varphi'<(1-4\mu)2\pi$ and $r\geq r_{b}=r_{0}\sin \theta_{M}$. So, this is just Minkowski spacetime with an angle deficit $2\alpha=8\pi\mu$.

Due to the fact that the junction surface $\mathcal{J}$ between the two metrics contains no shells or layers of matter, Israel conditions alluding to the continuity of the extrinsic curvature across the boundary surface reduce to the continuity of the metric and its normal derivatives on it. In the cylindrical coordinates adopted in (\ref{int2}) and (\ref{ext1}), this condition simply translates into
\begin{equation}\label{isra}
g_{\mu\nu}^{\,\,\,-}\mid_{\mathcal{J}}=g_{\mu\nu}^{\,\,\,+}\mid_{\mathcal{J}}\,,\,\,\,\,\,\,
\partial g_{\mu\nu}^{\,\,\,-}/\partial r\mid_{\mathcal{J}}=\partial g_{\mu\nu}^{\,\,\,+}/\partial r\mid_{\mathcal{J}}.
\end{equation}
From the interior point of view, the boundary surface $\mathcal{J}$ is defined by $r=r_{0}\theta_{M}$. The continuity of the metric on $\mathcal{J}$ then requires $r=r_{0}(1-4\mu)^{-1}\sin\theta_{M}$, as viewed from the exterior. Concerning the derivatives, we see that the only non trivial ones are
\begin{equation}\label{matchmenos}
\partial g_{\varphi\varphi}^{\,\,\,-}/\partial r\mid_{\mathcal{J}}=\partial g_{\varphi\varphi}^{\,\,\,+}/\partial r\mid_{\mathcal{J}}=2\,r_{0}\cos\theta_{M}\sin\theta_{M}.
\end{equation}
So, the value of $g_{\mu\nu}$ as well as $\partial g_{\mu\nu}/\partial r$ agree on both sides of the boundary, and the junction conditions are automatically satisfied. 

\begin{figure}
     \centering
     \begin{subfigure}[b]{0.25\textwidth}
         \centering
         \includegraphics[width=\textwidth]{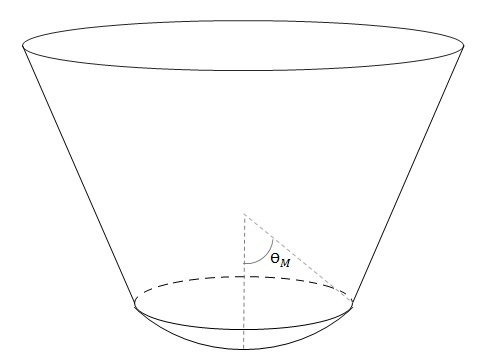}
         \caption{$\theta_{M}<\pi/2$, $0<\mu<1/4$}
         \label{fig1}
     \end{subfigure}
     \hfill
     
     \begin{subfigure}[b]{0.20\textwidth}
         \centering
         \includegraphics[width=\textwidth]{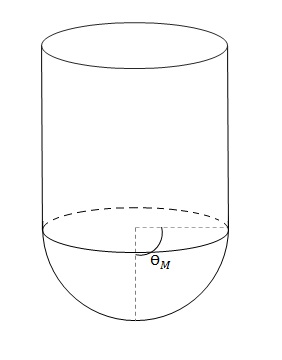}
         \caption{$\theta_{M}=\pi/2$, $\mu=1/4$}
         \label{fig2}
     \end{subfigure}
     \begin{subfigure}[b]{0.21\textwidth}
         \centering
         \includegraphics[width=\textwidth]{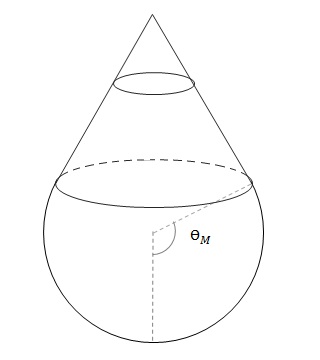}
         \caption{$\pi/2<\theta_{M}<\pi$, $1/4<\mu<1/2$}
         \label{fig3}
     \end{subfigure}
        \caption{Embedding diagrams for the cross sectional geometry corresponding to the interior and exterior solutions}
        \label{figs}
\end{figure}

The structure of the strings can be viewed in the embedding diagrams of Fig. \ref{figs}, where the cross sectional geometry $t=const$, $z=const$ is displayed. When $\theta_{M}<\pi/2$ (this is $0<\mu<1/4$), we have a spherical cap covering less than hemisphere for the interior metric, and a cone opening out and extending to infinite for the exterior metric. In turn, when $\theta_{M}=\pi/2$ ($\mu=1/4$), the geometry corresponds exactly to an hemisphere for the interior and a cylinder of radius $r_{0}$ for the exterior. Finally, if $\pi/2<\theta_{M}<\pi$ (thus, $1/4<\mu<1/2$), the interior is more than a hemisphere and the exterior solution is like a dunce cap sitting on top of the sphere.

Gott's construction makes use of the fact that the solution corresponding to two static parallel cosmic strings separated by a distance $2d$ is easily obtained from the one just discussed. We summarized here the findings made in \cite{Gott1}, primarily with the aim of setting the notation.  If we adopt cartesian coordinates
\begin{equation}\label{cambio}
x=r\cos(\varphi'+\alpha),\,\,\,\, y=r\sin(\varphi'+\alpha)+d,
\end{equation}
the exterior metric (\ref{ext2}) is just $ds^2=-dt^2+dx^2+dy^2+dz^2$, valid in the region $x^2+(y-d)^2\geq r_{b}$, and $\theta_{M}<\pi/2$, as in Fig. \ref{fig1}. Besides, we have to identify opposite points on the wedge, hence $(t,(y-d)\tan\alpha,y,z)$ and $(t,-(y-d)\tan\alpha,y,z)$ are really the same points. This corresponds to the string whose center is located at a distance $d$ measured along the $y$-axis. In the same way we can consider a mirror-image second copy located at $y=-d$; the two copies obey the matching conditions along  the three-surface $\mathcal{J}_{y}$ defined by $y=0$, which has metric $ds^2=-dt^2+dx^2+dz^2$ and thus zero intrinsic and extrinsic curvature.

The two string static solution just constructed has unnerving global properties. Referring to Fig. \ref{fig2}, we see that observer $B$ perceives three images of observer $A$ (they are both at rest with respect to the strings). In particular, if a light beam is emitted at $A$, it can travel directly to $B$ by following the null geodesic passing through the origin $O$, or by null geodesics passing through the events $E_{1}-E_{2}$ and $E_{3}-E_{4}$, both pairs being respectively identified. According to Fig. 2 we have
\begin{equation}
 w_{0}^2=(x_{0}-y_{0}\sin{\alpha})^2+(d+y_{0}\cos{\alpha})^2,
  \label{elomega}
\end{equation}
and the value of $y_{0}$ is chosen as to minimize $w_{0}$, i.e. $y_{0}=x_{0}\sin{\alpha}-d\cos{\alpha}$. Replacing this value into (\ref{elomega}) we obtain the simple expression
\begin{equation}
 w_{0}=x_{0}\cos{\alpha}+d\sin{\alpha}.
  \label{elomega2}
\end{equation}
However, the light beams going around the strings can beat the light beam passing through $O$ only if $w_{0}<x_{0}$, which is achieved provided $d<y_{0}$; the same is true for a rocket travelling at a high enough speed $\beta_{R}$ relative to the strings at rest. Concretely, let the rocket start its journey at $A$ (event $E_{i}$) and finish it at $B$ (event $E_{f}$) following the path through $E_{1}-E_{2}$, where
\begin{equation} \label{eliyelf}
E_{i}=\Big(-\beta_{R}^{-1}w_{0}, x_{0},0,0\Big),\,\,\,\,E_{f}=\Big(\beta_{R}^{-1}w_{0}, -x_{0},0,0\Big).\notag
\end{equation}
\begin{figure}[tbp]
\centering\includegraphics[width=0.5\textwidth]{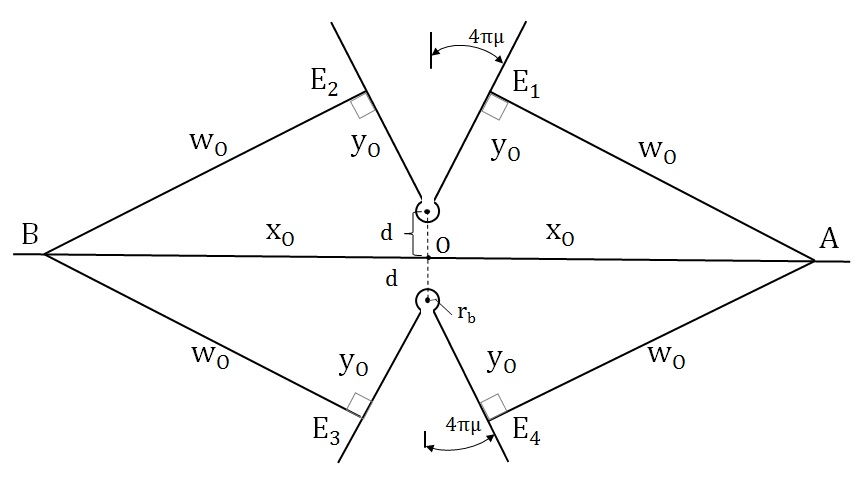}
\caption{Static, exterior two-string geometry for constant $z$.}
\label{fig2}
\end{figure}
Notice that the events $E_{i}$ and $E_{f}$ are spacelike separated only if $x_{0}^2-\beta_{R}^{-2}w_{0}^2>0$. Now, after performing a $+x$-boost with velocity $\beta_{s}$ in the $y\geq0$ region, the events $E_{i}$ and $E_{f}$ can be made simultaneous if $\beta_{s}=\beta_{R}^{-1}w_{0}x_{0}^{-1}$. The same can be done in the $y\leq0$ region by means of a $-x$-boost with the same speed $\beta_{s}$. In this way, the rocket is able to complete a CTC as viewed from the laboratory frame (this is the frame with respect to which the strings have equal but opposite momenta). This CTC is described by the timelike geodesic segments arising as a consequence of the travel from $E_{i}$ to $E_{f}$ through $E_{1}-E_{2}$, and back from $E_{f}$ to $E_{i}$ through $E_{3}-E_{4}$.

In Ref. \cite{Gott1} it was shown that the closed path just described is actually a CTC only if $\beta_{s}>w_{0}/x_{0}$. In view of Eq. (\ref{elomega2}), we have thus that a necessary and sufficient condition for the existence of such CTC is
\begin{equation} \label{nessuf}
\gamma_{s}=(1-\beta_{s}^2)^{-1/2}>(\sin{\alpha})^{-1}\,\,\, \Rightarrow\,\,\, \beta_{s}>\cos{\alpha}.
\end{equation}
Similar conclusions can be obtained by considering a different closed trajectory for the rocket, as the one described by the segments $E_{1}-E_{2}-E_{3}-E_{4}$. In this case, the rocket would undergo acceleration only at the identified events $E_{1}-E_{2}$ and $E_{3}-E_{4}$ (as opposed to the situation previously described, where the rocket undergoes acceleration only at $E_{i}$ and $E_{f}$). Worth of mention is the fact that, for a given $\beta_{s}>\cos{\alpha}$, all the pairs $A,B$ associated with sufficiently large values of $x_{0}$ are connected by a certain CTC. In other words, CTC's are not limited to exist in a compact region. Details on the global structure of Gott's spacetime can be found in Refs. \cite{Ori}.

\section{Gott's construction according to $f(T)$ gravity}\label{sec3}

\subsection{Infinitely long cosmic strings and the tetrad field}\label{sec3a}
In the context of the extended theories of gravity relying on absolute parallelism, the dynamical field is the tetrad $e^a(x^{\mu})$, and often it turns out confusing to think on the metric field $\bold{g}$ as the carrier of the gravitational degrees of freedom, even though the tetrad is required to be orthonormal in the (pseudo) Riemannian sense $\bold{g}=e^ae^b\eta_{ab}$. In the case of $f(T)$ gravity \cite{FF}, \cite{FB}, the equations of motion governing the full orientation of the tetrad are
\begin{eqnarray}
  \bigl[e^{-1}\partial_\mu(e\ S_a{}^{\mu\nu})+e_a^\lambda T^\rho{}_{\mu\lambda} S_\rho{}^{\mu\nu}\bigr]
  f^{\prime}+&\nonumber\\
  +\,S_a{}^{\mu\nu} \partial_\mu T f^{\prime \prime} - \frac{1}{4}
  e_a^\nu f =& -4\pi  e_a^\lambda T_\lambda{}^{\nu},
  \label{ecuaciones}
\end{eqnarray}
where an arbitrary (at least twice differentiable) function $f$ of the Weitzenb\"{o}ck pseudo-invariant $T= S^{a}{}_{\mu\nu} T_{a}{}^{\mu\nu}$ appears \cite{elt}, and primes denote differentiation with respect to $T$. Here $T^{a}_{\,\,\mu\nu}=\partial_{\nu}e^a_{\,\,\mu}-\partial_{\mu}e^a_{\,\,\nu}$, are the components of the \emph{torsion} coming from the Weitzenb\"{o}ck connection, and the components $S^{a}{}_{\mu\nu}$ of the \emph{superpotential} read
\begin{align*}
  S^{a}{}_{\mu\nu} = \frac{1}{4}
  (T^{a}{}_{\mu\nu} - T_{\mu \nu }{}^{a} + T_{\nu \mu }{}^{a}) +
  \frac{1}{2} (\delta_{\mu}^{a} T_{\sigma\nu}{}^{\sigma} -
  \delta_{\nu}^{a} T_{\sigma\mu}{}^{\sigma})\,,
\end{align*}
where $T_{\sigma }{}^{\sigma\nu}=g^{\mu\rho}\,T_{\mu\rho }{}^{\nu}=e^{\mu}_{\,\,a}e^{\rho}_{\,\,b}\,\eta^{ab}\,T_{\mu\rho }{}^{\nu}$. Let us notice that the tetrad components $e^a_{\,\,\mu}$ carry Lorentz-type indexes $a$ as well as spacetime ones $\mu$ . The equations (\ref{ecuaciones}) are derived from the action
\begin{align}
  I=\frac{1}{16 \pi  }\int f(T)\,  e\, d^4x+I_{\rm matter}\,,
  \label{actionfT3D0}
\end{align}%
which reduces to GR (in its teleparallel equivalent form), when $f(T)=T$. The energy-momentum tensor $T_\lambda{}^{\nu}$ appearing in the RHS of the equations is derived from $I_{\rm matter}$ in the usual way. For a bibliographical compendium on $f(T)$ gravity, the reader is invited to consult \cite{Review}.

We proceed now to discuss both string (interior and exterior) solutions within this context. The interior tetrad is just
\begin{equation}\label{inttet}
e^0_{-}=dt,\,\,\,\,e^1_{-}=dr,\,\,\,\,e^2_{-}=r_{0}\sin(r/r_{0})d\varphi,\,\,\,\,e^3=dz\,
\end{equation}
which corresponds to the form (\ref{int2}) of the interior line element. Due to the simplicity of the tetrad field, just a few components of the torsion and contortion tensor are non trivial. In spacetime, totally covariant form, they are
\begin{equation}\label{int00}
T_{\varphi\, r\,\varphi}=\frac{-r_{0}}{2}\sin{(2r/r_{0})},\,\,\,\,S_{t\,t\,r}=S_{z\,r\,z}=\frac{\cot{(r/r_{0})}}{2r_{0}}.\notag
\end{equation}
It is straightforward to check from these that $T=0$. This leaves us with only two non trivial motion equations, namely
\begin{equation}\label{int01}
f+2f'/r_{0}^2=16 \pi \, T_t{}^{t}, \,\,\,\,\,\,\,\, f+2f'/r_{0}^2=16 \pi \, T_z{}^{z},
\end{equation}
then we obviously have $T_t{}^{t}=T_z{}^{z}$, which implies $\rho_{0}=-p_{z}$, as we mentioned before. Due to the null character of $T$, we see that $f(0)=0$ and $f'(0)=1$ for any smooth function $f$ connected to the identity, i.e. $f(T)=T+\mathcal{O}(T^2)$. In this way, the equations reduce to the condition $\rho_{0}=-p_{z}=1/8 \pi r_{0}^2$. This proves that the tetrad (\ref{inttet}) is a solution of the $f(T)$ field equations for an energy momentum tensor $T_\mu{}^{\nu}=diag(\rho_{0},0,0,-p_{z})$ verifying $\rho_{0}=-p_{z}=1/8 \pi r_{0}^2$, for any smooth deformation of GR.

For the exterior tetrad we consider again the "square root'' of (\ref{ext1}), then we have
\begin{equation}\label{exttet}
e^0_{+}=dt,\,\,\,\,e^1_{+}=dr,\,\,\,\,e^2_{+}=(1-4\mu)r\,d\varphi,\,\,\,\,e^3_{+}=dz\,,
\end{equation}
which leads us to
\begin{equation}\label{int02}
T_{\varphi\, r\,\varphi}=-r\,(1-4\mu)^2,\,\,\,\,\,\,\,S_{t\,t\,r}=S_{z\,r\,z}=1/2r.\notag
\end{equation}
This tetrad also conduces to $T=0$ and, then, it automatically solves the vacuum $f(T)$ field equations. The fact that (\ref{inttet}) and (\ref{exttet}), being GR's solutions, remain as solutions of the $f(T)$ equations, it is ultimately a consequence of both having $T=0$; this means that (\ref{ecuaciones}) reduces to
\begin{eqnarray}
  e^{-1}\partial_\mu(e\ S_a{}^{\mu\nu})+e_a^\lambda T^\rho{}_{\mu\lambda} S_\rho{}^{\mu\nu} = -4\pi  e_a^\lambda T_\lambda{}^{\nu},
  \label{ecuacionesterg}
\end{eqnarray}
provided $f(T)=T+\mathcal{O}(T^2)$; these are no other than the TEGR equations with $T=0$ ($\partial_\mu T$ vanishes). Therefore, any solution $\bold{g}=e^a e^b\eta_{ab}$ of GR having tetrads $e^a$ leading to $T=0$, remains as solution of $f(T)$ gravity. Diagonal, proper tetrads as (\ref{inttet}) and (\ref{exttet}) are rare among solutions of $f(T)$ gravity. They only work in very simple and symmetric situations, as the one corresponding to the constant curvature spaces here considered. The selection of preferred frames is a distinct feature of $f(T)$ gravity in its \emph{pure tetrad} formulation, opposed (only in nature), to the \emph{covariant} approach developed in \cite{covfor}, see the appeasing discussion of Ref. \cite{grupo23} in regard with the two approaches.

The tetrads (\ref{inttet}) and (\ref{exttet}) $\mathcal{C}^{1}$-match on $\mathcal{J}$. As a matter of fact, the continuity of the tetrad and its first derivative on the matching surface follow at once from
\begin{eqnarray}\label{matchtet}
&&e^2_{-}\mid_{\mathcal{J}}=e^2_{+}\mid_{\mathcal{J}}=r_{0}\sin\theta_{M}, \notag\\
&&\partial e^2_{-}/\partial r\mid_{\mathcal{J}}=\partial e^2_{+}/\partial r\mid_{\mathcal{J}}=\cos \theta_{M},\notag
\end{eqnarray}
being the remaining components trivially matched. This fact ensures that the exterior and interior tetrads together constitute a well defined global parallelization, see Refs. \cite{junction1} for further discussions on this subject.

\subsection{Gott's construction and remnant symmetries}\label{sec3b}

Once the general structure of the string is presented, we focus on the exterior, vacuum tetrad (we drop the subscripts $+$ and $-$ from know on). Because we are planning to boost the strings in the $x$-direction (with opposite speeds), it is convenient to write the tetrad in cartesian coordinates, defined by (\ref{cambio}). The tetrad (\ref{exttet}) in these coordinates looks 
\begin{eqnarray}\label{exttetcart}
e^0_{\uparrow}&=&dt,\,\,\,\,\,\,\,\,
e^1_{\uparrow}=\frac{x\,dx}{r_{{\uparrow}}}+\frac{(y-d)\, dy}{r_{{\uparrow}}},\notag\\
e^2_{\uparrow}&=&-\frac{(y-d)\, dx}{r_{{\uparrow}}}+\frac{x\,dy}{r_{{\uparrow}}},\,\,\,\,\,\,\,\,e^3_{\uparrow}=dz,
\end{eqnarray}
which is valid in the region $y\geq0$ corresponding to the string whose center is at $y=d$, and we wrote $r_{\uparrow}^2=x^2+(y-d)^2$. A similar coordinate change is used to obtain the second copy valid in $y\leq0$ (this implies to change $y$ by $-y$ in $(\ref{exttetcart}$)), which reads 
\begin{eqnarray}\label{exttetcart2}
e^0_{\downarrow}&=&dt,\,\,\,\,\,\,\,\,
e^1_{\downarrow}=\frac{x\,dx}{r_{\downarrow}}+\frac{(y+d)\, dy}{r_{\downarrow}},\notag\\
e^2_{\downarrow}&=&\frac{(y+d)\, dx}{r_{\downarrow}}-\frac{x\,dy}{r_{\downarrow}},\,\,\,\,\,\,\,\,e^3_{\downarrow}=dz,
\end{eqnarray}
where now $r_{\downarrow}^2=x^2+(y+d)^2$. Let us remark that both tetrads are continuous and differentiable on $\mathcal{J}_{y}$.

In order to put the strings in motion we now apply coordinate-dependent boosts in $\pm x$. For the moment, we are not forced to consider the same (opposite) speeds for both strings, so we will have after the boosts
\begin{eqnarray}\label{exttetcartboostm}
e^{0'}_{\uparrow}&=&\cosh[\phi_{\uparrow}]dt-\frac{x\,\sinh[\phi_{\uparrow}]}{r_{\uparrow}}\,dx-\frac{(y-d)\,\sinh[\phi_{\uparrow}]}{r_{\uparrow}}\,dy\notag\\
e^{1'}_{\uparrow}&=&-\sinh[\phi_{\uparrow}]dt+\frac{x\,\cosh[\phi_{\uparrow}]}{r_{\uparrow}}\,dx+\frac{(y-d)\,\cosh[\phi_{\uparrow}]}{r_{\uparrow}}\,dy,\notag\\
e^{2'}_{\uparrow}&=&-\frac{(y-d)\, dx}{r_{\uparrow}}+\frac{x\,dy}{r_{\uparrow}},\,\,\,\,\,\,\,\, e^{3'}_{\uparrow}=dz,
\end{eqnarray}
\begin{eqnarray}\label{exttetcartboostne}
e^{0'}_{\downarrow}&=&\cosh[\phi_{\downarrow}]dt+\frac{x\,\sinh[\phi_{\downarrow}]}{r_{\downarrow}}\,dx+\frac{(y+d)\,\sinh[\phi_{\downarrow}]}{r_{\downarrow}}\,dy\notag\\
e^{1'}_{\downarrow}&=&\sinh[\phi_{\downarrow}]dt+\frac{x\,\cosh[\phi_{\downarrow}]}{r_{\downarrow}}\,dx+\frac{(y+d)\,\cosh[\phi_{\downarrow}]}{r_{\downarrow}}\,dy,\notag\\
e^{2'}_{\downarrow}&=&\frac{(y+d)\, dx}{r_{\downarrow}}-\frac{x\,dy}{r_{\downarrow}},\,\,\,\,\,\,\,\,e^{3'}_{\downarrow}=dz,
\end{eqnarray}
for $y\geq0$ and $y\leq0$ regions, respectively. In both boosted tetrads we have $\phi_{\uparrow\downarrow}=\phi_{\uparrow\downarrow}(t,\bar{x})$ (see Ref. \cite{Andronikos} for a full analysis concerning the role of the \emph{booston} $\phi$ in 2D toy models). Notice that (\ref{exttetcartboostm}) and (\ref{exttetcartboostne}) have opposite rapidities $\phi_{\uparrow\downarrow}=\tanh^{-1}[\beta_{\uparrow\downarrow}]$, so they are boosts in $x$ with opposite speeds. Gott's construction discussed in section \ref{sec2} requires $\phi_{\uparrow}=\phi_{\downarrow}=\tanh^{-1}[\beta_{s}]=constant$.

Due to the fact that we locally boosted solutions of a theory which is not local Lorentz invariant in general, two questions naturally arise:


(a) Under what condition, if any, are (\ref{exttetcartboostm}) and (\ref{exttetcartboostne}) solutions of the $f(T)$-motion equations?

(b) Once we established that $e^{a'}_{\uparrow}$ and $e^{a'}_{\downarrow}$ are solutions, is that true for $e^{a'}_{\uparrow}\cup e^{a'}_{\downarrow}$? If that were to be true, it would entail the fulfillment of the junction conditions on $\mathcal{J}_{y}$.

The answer to the first question resides in the functional form of the Weitzenb\"{o}ck pseudo-invariant $T$, which clearly is not invariant under local Lorentz transformations in the tangent space. After some work it can be obtained that 
\begin{equation}\label{Ttransf}
T_{\uparrow\downarrow}=\mp\,2\,\partial_{t}\phi_{\uparrow\downarrow}\,/\,r_{\uparrow\downarrow},
\end{equation}
where, from now on, upper signs corresponds to $\uparrow$, and lower ones to $\downarrow$. This means that the remnant group of transformations associated with the original tetrads (\ref{exttetcart}) and (\ref{exttetcart2}), i.e. the local Lorentz transformations leaving invariant the null value of $T$ \cite{grupo}, includes --among many others-- position-dependent boosts in the $t-x$ plane. In other words, (\ref{exttetcartboostm}) and (\ref{exttetcartboostne}) are solutions of the vacuum $f(T)$ field equations only if $\phi_{\uparrow\downarrow}=\phi_{\uparrow\downarrow}(\bar{x})$ alone.

Once we established that $\phi_{\uparrow\downarrow}\neq\phi_{\uparrow\downarrow}(t)$, we proceed to answer question (b) raised above. It is quite evident that on the junction surface $\mathcal{J}_{y}$ ($y=0$) the tetrad is not $\mathcal{C}^1$; actually it is not even continuous due to the jump in the sign. This is so provided $\phi_{\uparrow\downarrow}|_{0}\doteq\phi_{\uparrow\downarrow}(x,y=0,z)\neq0$. There is also a discontinuity of the derivatives of $\phi_{\uparrow\downarrow}$ on $\mathcal{J}_{y}$, as witnessed by the presence of several non-null components of the torsion, e.g  
\begin{equation}\label{nnull}
T^{\uparrow\downarrow}_{x\,t\,x_{i}}=\mp\frac{x\,\partial_{x_{i}}\phi_{\uparrow\downarrow}}{r_{\uparrow\downarrow}},\,\,\,\,\,\,\,\,\,T^{\uparrow\downarrow}_{y\,t\,x_{i}}=\pm\frac{(d\mp y)\,\partial_{x_{i}}\phi_{\uparrow\downarrow}}{r_{\uparrow\downarrow}}, 
\end{equation}
where $x_{i}:x,y,z$. Other components of the torsion that do not involve $\phi_{\uparrow\downarrow}$ exist as well, for instance
\begin{equation}\label{nnullsin}
T^{\uparrow\downarrow}_{x\,y\,x}=(d\mp y)\,/\,r_{\uparrow\downarrow},\,\,\,\,\,\,\,\,\,\,T^{\uparrow\downarrow}_{y\,x\,y}=x\,/\,r_{\uparrow\downarrow}, 
\end{equation}
but they are automatically continuous on $\mathcal{J}_{y}$, because $r_{\uparrow}=r_{\downarrow}$ there. Hence, in order to have a well defined global parallelization, we need to demand $\phi_{\uparrow\downarrow}|_{0}=0$ and the continuity of the  derivatives across $\mathcal{J}_{y}$. This simple conclusion puts several constrains on the functional form of $\phi_{\uparrow\downarrow}$ near $\mathcal{J}_{y}$; in particular, it rules out the case $\phi_{\uparrow\downarrow}=constant$ lying behind Gott's construction.

\bigskip

A crucial aspect in Gott's construction is the fact that the events $E_{i}$ and $E_{f}$, both belonging to the matching surface defined by $y=0$, become simultaneous in the laboratory frame after the action of the boosts in both directions of the $x$-axis; this simply requires $\beta_{s}=w_{0} \beta_{R}^{-1}x_{0}^{-1}$, which implies $\phi(\bar{x})\neq0$. In contrast, $f(T)$ motion equations demand $\phi=0$ ($\beta_{s}=0$) along the matching surface, thus selecting frames (tetrads) which necessarily undergo accelerated motion, at least within a strip $-\epsilon_{\downarrow}<y<\epsilon_{\uparrow}$, for non null, small $\epsilon_{\downarrow}$ and $\epsilon_{\uparrow}$. 

In effect, in the original static solution for $y\geq0$, let us consider the events $E_{i}^{\uparrow}$ and $E_{f}^{\uparrow}$  described by
\begin{equation}\label{nevents}
E_{i}^{\uparrow}=\Big(-\beta_{\uparrow}^{-1}w_{\uparrow}, x_{0},\epsilon_{\uparrow},0\Big),\,\,\,\,E_{f}^{\uparrow}=\Big(\beta_{\uparrow}^{-1}w_{\uparrow}, -x_{0},\epsilon_{\uparrow},0\Big),\notag
\end{equation}
where now $w_{\uparrow}=x_{0}\cos{\alpha}+(d-\epsilon_{\uparrow})\sin{\alpha}$ (see Eq. (\ref{elomega2})), and $\beta_{\uparrow}$ is the constant speed of the rocket in $y\geq\epsilon_{\uparrow}$. Again, the separation of $E_{i}^{\uparrow}$ and $E_{f}^{\uparrow}$ is spacelike provided $x_{0}^2-\beta_{\uparrow}^{-2}w_{\uparrow}^2>0$, and they become simultaneous in the laboratory frame after the action of a $+x$-boost with speed $\beta_{s\uparrow}=w_{\uparrow}\, \beta_{\uparrow}^{-1}x_{0}^{-1}$. As $\beta_{\uparrow}<1$, we have then $\beta_{s\uparrow}>w_{\uparrow}/x_{0}$, or
\begin{equation}\label{ndoblev2}
\beta_{s\uparrow}>\cos{\alpha}+(d-\epsilon_{\uparrow})\sin{\alpha}/x_{0}.
\end{equation}
However, relation (\ref{ndoblev2}) is not consistent with the fact that $\beta_{s\uparrow}=\tanh{\phi_{\uparrow}}$ must go to $0$ as $\epsilon_{\uparrow}\rightarrow 0$; this simply means that $E_{i}^{\uparrow}$ and $E_{f}^{\uparrow}$ cannot be simultaneous as $\epsilon_{\uparrow}\rightarrow 0$. This same conclusion applies to the corresponding events in $y\leq0$, which in the static picture have coordinates 
\begin{equation}\label{neventsbajo}
E_{i}^{\downarrow}=\Big(-\beta_{\downarrow}^{-1}w_{\downarrow},- x_{0},-\epsilon_{\downarrow},0\Big),\,\,  
E_{f}^{\downarrow}=\Big(\beta_{\downarrow}^{-1}w_{\downarrow}, x_{0},-\epsilon_{\downarrow},0\Big),\notag
\end{equation}
where now $\beta_{\downarrow}$ is the constant speed of the rocket in $y\leq\-\epsilon_{\downarrow}$ (perhaps different from $\beta_{\uparrow}$), and $w_{\downarrow}=x_{0}\cos{\alpha}+(d-\epsilon_{\downarrow})\sin{\alpha}$. These events can be made simultaneous if a boost in the $-x$-direction with speed $\beta_{s\downarrow}=w_{\downarrow}\, \beta_{\downarrow}^{-1}x_{0}^{-1}$ is performed, but \emph{only} if $\epsilon_{\downarrow}$ is not too small. 

In this way, one can conceive a rocket travelling at a high enough speed along a (spatially) closed curve defined by the timelike geodesic segments $E_{i}^{\uparrow}-E_{f}^{\uparrow}$ (through $E_{1}-E_{2}$) and $E_{i}^{\downarrow}-E_{f}^{\downarrow}$ (through $E_{3}-E_{4}$), plus two non-geodesic timelike segments $E_{f}^{\uparrow}-E_{i}^{\downarrow}$ and $E_{f}^{\downarrow}-E_{i}^{\uparrow}$. Once the two strings are in motion with speeds $\beta_{s\uparrow}$ and $\beta_{s\downarrow}$ in opposite directions, the rocket departs from the position corresponding to $E_{i}^{\uparrow}$ and arrives simultaneously to the one of $E_{f}^{\uparrow}$ in the laboratory frame. Then it takes the rocket a time $\tau$, as measured in the laboratory frame, to cover the non geodesic segment from $E_{f}^{\uparrow}$ to $E_{i}^{\downarrow}$; during this stage the rocket experiences a strong accelerated motion due to the fact that it must dramatically change its speed because the boosts $\beta_{s}=\tanh{\phi}$ go to zero as $y\rightarrow 0$. Immediately after that, the rocket is able to travel again instantaneously between the positions corresponding to $E_{i}^{\downarrow}$ and $E_{f}^{\downarrow}$ by using the string moving along $-x$ in $y<0$. Finally, it will take another time $\tau$ to go from $E_{f}^{\downarrow}$ to its original starting position. The rocket thus will complete a closed spatial circuit in a time $2\tau$ according to the laboratory frame. This circuit do not constitute a closed timelike curve. Even though $\tau$ could be very small --this depends on the (unspecified) state of motion along the accelerated stages-- the rocket cannot travel back in time to its own past, although it might \emph{almost} do it.

Things are not much better for the rocket following the closed path $E_{1}-E_{2}-E_{3}-E_{4}$, where, again, $E_{1}-E_{2}$ and $E_{3}-E_{4}$ are identified. After the Lorentz transforms have been made in the region $\mid y\mid \geq \epsilon$ (we are considering now $\epsilon=\epsilon_{\downarrow}=\epsilon_{\uparrow}$), the events $E_{2}-E_{3}$ and $E_{4}-E_{1}$ are simultaneous in the laboratory frame, just as before. However, the rocket cannot travel from $E_{2}$ to $E_{3}$ and from $E_{4}$ to $E_{1}$ without crossing the region $\mid y\mid \leq \epsilon$, where $\beta_{s}$ must change its value in order to comply with the matching conditions. More precisely, let us consider the following events in the static frame $(t,\bar{x})$
\begin{eqnarray}\label{mirrore}
&&E_{1}^{\prime}=(0,x_{1},\epsilon,0),\,\,\,\,\,\,\,\,\,\,\,\,\, E_{2}^{\prime}=(0,-x_{1},\epsilon,0),\notag\\ 
&&E_{3}^{\prime}=(0,-x_{1},-\epsilon,0),\,\,\,\, E_{4}^{\prime}=(0,x_{1},-\epsilon,0).\notag
\end{eqnarray}
These events are no other than the projection of $E_{1};E_{2};E_{3};E_{4}$ onto the boundary of the accelerated strip. After the corresponding boosts with (constant) opposite speeds $\beta_{s}$ are applied in the $y\geq \epsilon$ and $y\leq \epsilon$ regions, these events have coordinates in the laboratory frame 
\begin{align}\label{mirrore2}
&E_{1}^{\prime}=(\gamma_{s} \beta_{s}x_{1},\gamma_{s} x_{1},\epsilon,0),\,\,\, E_{2}^{\prime}=(-\gamma_{s} \beta_{s}x_{1},-\gamma_{s} x_{1},\epsilon,0),\notag\\
&E_{3}^{\prime}=(\gamma_{s} \beta_{s}x_{1},-\gamma_{s} x_{1},-\epsilon,0),\,E_{4}^{\prime}=(-\gamma_{s} \beta_{s}x_{1},\gamma_{s} x_{1},-\epsilon,0).\notag
\end{align}
Even though $E_{1}^{\prime}$ and $E_{3}^{\prime}$ as well as $E_{2}^{\prime}$ and $E_{4}^{\prime}$ are simultaneous in the laboratory frame, a rocket at speed $\beta_{r}$ in the same frame will cover the two segments $E_{2}-E_{2}^{\prime}$ and $E_{3}^{\prime}-E_{3}$ only if
\begin{equation}\label{condro}
\beta_{r} \beta_{s} \gamma_{s} \sin{\alpha} = (d-\epsilon)/y_{0}+\cos{\alpha}.
\end{equation}
Again, we always can choose $x_{0}$ in such a way that $y_{0}>>d-\epsilon$, and, since $\beta_{r}<1$ we end up with $\gamma_{s}>(\sin{\alpha})^{-1}$ as before. But we know for certain that $\beta_{s}$ goes to zero as $\epsilon \rightarrow 0$, so the last inequality cannot be true in that limit. This simply shows that it is not possible to travel by rocket at speed $\beta_{r}<1$ in the laboratory frame with the aim of joining the simultaneous events $E_{1}^{\prime}$ and $E_{3}^{\prime}$; in other words, it will take a time (let us say, again) $\tau$ to cover the segment $E_{1}^{\prime}$ and $E_{3}^{\prime}$ from the point of view of an observer in the laboratory frame. Obviously, the same conclusion applies to the way back from $E_{3}^{\prime}$ to $E_{1}^{\prime}$. The rocket will spend another time $\tau$ to cover that trip, and then, a time $2\tau$ to circle the two strings.  
\section{Final comments}
We have seen that the space traveler --having the intention to become a time traveler as well--will spent, roughly, a time $2\tau$ to circumnavigate the strings in rapid opposite motion, according to an observer located at the laboratory frame. At the moment the value of $\tau$ is uncertain, because it depends on the kinematics unfolding within the accelerated strip, but we do know that $\tau\neq0$. Causal pathologies then might be on the verge to occur due to the fact that the traveller still could be in the position to influence his/her own past because $\tau$ could be very small. It would be interesting, then, to estimate $\tau$ by analysing different motions obeying the constraint $\beta_{s}=0$ on $\mathcal{J}_{y}$, and to inquire if the results here obtained are present as well in other theories of gravity relying on absolute parallelism as, for instance, in Born-Infeld gravity \cite{Vatu}.

In the original construction of Ref. \cite{Gott1} it was mentioned that the strings need not be parallel. In the present context, it would be relevant to figure out what kind of rotations mapping the three-surface $\mathcal{J}_{y}$ into itself also belong to the remnant group associated to the tetrads (\ref{exttetcart}) and (\ref{exttetcart2}), with the aim to collect, thus, further arguments against the formation of CTC's in Gott's space. 

\bigskip

\begin{center}
   \large $\ast \; \ast \; \ast$   
\end{center}

\emph{Acknowledgements}. I would like to express my gratitude to Diego Mazzitelli for reading the manuscript and making valuable comments. The author is member of \emph{Carrera del Investigador Cient\'{i}fico} (CONICET), and his work is supported by CONICET and Instituto Balseiro (UNCUYO).

\end{document}